\documentclass{PoS}

\usepackage[authoryear,square]{natbib}
\bibpunct{(}{)}{;}{a}{}{,}

\let\ifpdf\relax

\title{The SKA view of the Neutral Interstellar Medium in Galaxies}

\ShortTitle{Neutral ISM in Galaxies}

\author{\speaker{W.J.G.~de~Blok}$\,^{1,2,3}$,  F.~Fraternali$^{4,3}$, G.H.~Heald$^{1,3}$, E.A.K.~Adams$^{1}$, A.~Bosma$^5$, B\"arbel~S.~Koribalski$^6$
and the \HI\ Science Working Group\\
\llap{$^1$}ASTRON\\
           Postbus 2, 7990 AA, Dwingeloo, The Netherlands\\
\llap{$^2$}Department of Astronomy, Astrophysics, Cosmology and Gravity Centre, \\
          University of Cape Town, Private Bag X3, Rondebosch, 7701 South Africa\\
\llap{$^3$}Kapteyn Astronomical Institute, University of Groningen\\
          Postbus 800, 9700 AV, Groningen, The Netherlands\\
\llap{$^4$}Department of Physics and Astronomy, University of Bologna,\\ 
           via Berti Pichat 6/2, I-40127 Bologna, Italy\\
\llap{$^5$}Aix Marseille Universite, CNRS, LAM (Laboratoire d'Astrophysique de Marseille)\\ 
           UMR 7326, F-13388 Marseille, France\\
\llap{$^6$}Australia Telescope National Facility, CSIRO Astronomy \& Space Science\\
           P.O. Box 76, Epping, NSW 1710, Australia\\
E-mail: \email{blok@astron.nl}}

\abstract{Two major questions in galaxy evolution are how
  star-formation on small scales leads to global scaling laws and how
  galaxies acquire sufficient gas to sustain their star formation
  rates.  \HI\ observations with high angular resolution and with
  sensitivity to very low column densities are some of the important
  observational ingredients that are currently still missing.  Answers
  to these questions are necessary for a correct interpretation of
  observations of galaxy evolution in the high-redshift universe and
  will provide crucial input for the sub-grid physics in
  hydrodynamical simulations of galaxy evolutions. In this chapter we
  discuss the progress that will be made with the SKA using targeted
  observations of nearby individual disk and dwarf galaxies.}

\FullConference{
Advancing Astrophysics with the Square Kilometre Array\\
June 8-13, 2014\\
Giardini Naxos, Italy}

\newcommand{\HI}{{\sc H\,i}}

\begin{document}

\section{The Galactic Gas Cycle}

Galaxy evolution is driven by the flow of gas into galaxies, the
transformation of gas into stars, and the expulsion of gas due to the
subsequent stellar evolution processes.  Atomic neutral hydrogen (\HI)
is an excellent tracer --- and often the main constituent --- of this
gas, and can be observed readily in the 21-cm line.  The SKA will be
able to directly trace the gradual transformation from primordial
hydrogen into galaxies over cosmic time. However, to correctly
interpret this evolution, direct detailed observations of the
sub-kpc-scale physical processes that cause this transformation are
essential for understanding the astrophysics of galaxy evolution.

Two very important processes in this cosmic evolution are star
formation and gas accretion. The first one enriches the interstellar
medium (ISM) with metals, injects energy into it, and, through the
ejection of gas, also enriches the intergalactic medium (IGM). The
second process delivers gas from outside galaxies (either primordial
or previously ejected) into the star-forming disks, which ensures
galaxies can keep forming stars over a Hubble time.

Star formation thus plays a central role in galaxy evolution, yet
little is known about the necessary conditions for star formation to
occur. Global, kpc-scale relations, such as the Kennicutt-Schmidt
Law \citep{kenn89, Kennicutt1998}, have established that there is a
direct connection between the (molecular) gas surface density and the
star formation rate (SFR) surface density. This relation has been
found to hold for both spiral galaxies, as well as late-type dwarf
galaxies (see, e.g., \citealt{Leroy2008} and \citealt{Bigiel2008} for
a recent analysis). How the underlying astrophysics at the scales of
individual gas clouds and cloud complexes leads to such a relation on
more global scales remains, however, ill-understood.

The initial phase of this star-formation sequence will presumably be
the cooling of gas to temperatures below $\sim 10^4$ K, where it can
collapse and form clouds.  Observing this cold, clumpy ISM and
distinguishing it from the warmer, more diffuse phase requires imaging
at high spatial and velocity resolution. To date, few
\HI\ observations of nearby galaxies resolve the individual cold gas
complexes (see, e.g., \citealt{braun2009} and \citealt{kim2003} for
examples in the Local Group).  The SKA will have sufficient
sensitivity at the required high resolution to do this for many
galaxies. This enables studies of the properties of the cold gas in
galaxies at the desired cloud-size resolution over a wide range of
galaxy properties and environments. In addition, ALMA can observe the
molecular components of these clouds and complexes, and this
complementarity will lead to a revolutionary improvement in our
understanding of star formation --- and the conditions leading to it
--- in different environments.

A related question is: how can galaxies sustain their star formation
over a Hubble time? In general, local galaxies only have enough gas to
sustain their SFR for few Gyr, and they must thus acquire gas from
somewhere else (see \citealt{sancisi} for a recent
overview). Numerical simulations suggest gas flows in from the IGM (or
``cosmic web''), through a process called ``cold accretion'' (e.g.,
\citealt{keres}). ``Cold'' in this context means that the gas has not
been shock-heated as it entered the galaxy halo. This cold accretion
process is discussed in more detail in the chapter by
\citet{popping2014} elsewhere in this volume.  So far there is little
direct observational evidence for significant cold accretion. The
observed cold gas accretion in galaxies seems to be an order of
magnitude too low to explain the current star formation rates (SFR) in
galaxies \citep{sancisi,putman}. If cold accretion is the dominant
process by which galaxies acquire their gas, then current
observational limits indicate it must happen at \HI\ column densities
below $\sim 10^{18}$ cm$^{-2}$.  Only a limited number of galaxies
have so far been explored at sufficient sensitivity and spatial
resolution to probe this regime \citep{heald11, putman}. SKA will be
able to directly detect and map in detail the gas at these column
densities and trace its connection with the cosmic web.

The galactic fountain process \citep{shapiro, bregman, norman} links
the gas in the star forming disk and that in the halo.  Massive stars,
through supernova explosions and stellar winds, can push gas out of
the disk and into the halo of a galaxy. This creates the holes and
bubbles frequently observed in the gas disks of galaxies
\citep{bagetakos}.  The expelled gas will cool and eventually rain
back on the disk, most likely in the form of \HI\ clouds
\citep{putman}.  Such clouds have also been observed in a number of
other galaxies as part of an extra-planar gas component
\citep{sancisi}, and presumably form the equivalent of the high and
intermediate velocity clouds (HVCs and IVCs, respectively) in our
Galaxy.  It is thought that the process of these clouds moving through
the hot gaseous halo of a galaxy provides an alternative mechanism for
accretion of gas.  Here, hot halo gas cools in a cloud's wake and is dragged
along as the cloud moves back into the disk \citep{ff13}.

While the galactic fountain clouds thus provide a possible means for
the transportation of gas into the disk, they also hinder our ability
to identify primordial gas clouds that are being accreted from the IGM.
A small number can be identified as they are counter-rotating with
respect to their host galaxy \citep{oosterloo}, but a better census
down to lower \HI\ masses and column densities is needed to properly
understand these accretion mechanisms. Prime targets for this are
galaxies in the `Local Volume', defined as the sphere of radius 10~Mpc
centered on the Local Group \citep{kar04,kar13,koribalski}. This
volume includes around 900 galaxies, the majority of which are
gas-rich. Distances to these galaxies are known with high accuracy.
The \HI\ contents and sizes of Local Volume galaxies cover more than
two orders of magnitudes, ranging from low-mass dwarf galaxies with
diameters of less than one kpc (e.g., Leo~T; \citealt{ryanweber}) to
grand design spirals with \HI\ diameters of $\sim$100~kpc (e.g.,
Circinus; \citealt{for}).

In summary, the SKA will be a unique instrument that will be able to
track the gas in low-redshift galaxies as it accretes onto the disk,
forms \HI\ cloud complexes (which in turn produce stars through an
intermediate, molecular phase), is expelled through feedback
processes, and subsequently returns to the disk. Complemented by ALMA
molecular line observations, the transformation from neutral atomic
and molecular gas to the formation of stars can thus be investigated
in exceptional detail over a large range in galaxy properties (see,
e.g., \citealt{ott2008} and \citealt{fukui} for work on the LMC and \citealt{stanimirovic}
for our Galaxy). These processes can be studied in detail in a variety
of environments at low redshift, thus establishing the astrophysical
foundations that are needed to interpret observations of galaxies at
higher redshifts, and to guide the implementation of `sub-grid'
physics into hydrodynamical simulations of galaxy formation and
evolution.

In the rest of this section we calculate the \HI\ column density and
mass sensitivities of SKA1-MID, SKA1-SUR and SKA2 for a number of
observing scenarios. This chapter specifically deals with
high-resolution and high-sensitivity observations of individual
objects, i.e., deep observations of limited areas of sky.  

The remainder of this chapter then discusses the prospects for
obtaining detailed, sensitive observations of nearby, individual
galaxies and the progress that will be made towards a better
understanding of the link between star formation, accretion, and gas
in these galaxies.

\subsection{SKA column density and \HI\ mass sensitivities\label{obssec}}

We calculate column density sensitivities for SKA1-MID,
SKA1-SUR and SKA2 for a range of angular and velocity resolutions and three
representative integration times. For the latter we choose:

$\bullet$ {\bf 10 hours}: representing a typical ``single-track''
observation (as motivated in Sect.\ \ref{sec:sec2_3}). In terms of science, an
observation like this can be used to characterise high-resolution
structures in the star forming disks of galaxies, as well as (at lower
spatial resolutions) the outer low-column density disk down to $N_{\rm
  HI} \sim 10^{18}$ cm$^{-2}$;

$\bullet$ {\bf 100 hours}: this longer observation will typically
result in detection of \HI\ column densities down to $\sim 10^{17}$
cm$^{-2}$ where accretion and faint outer gas can be directly studied;

$\bullet$ {\bf 1000 hours}: this extremely long observation allows
detection of column densities well below $10^{17}$ cm$^{-2}$ where one can
directly explore the link of galaxies with the cosmic web.

For the angular resolutions we choose representative beam sizes
between $1''$ and $30''$. For the velocity resolution we choose values
between $1$ and $20$ km s$^{-1}$.  Using these parameters we calculate
the $5\sigma$ column density detection limits per velocity channel.
Our basis for this are the SKA1-MID and SKA1-SUR natural sensitivities
as listed in Table 1 of the SKA Baseline Design Document
\citep{BD}. For SKA1-MID we use the sensitivity listed for 190 SKA
dishes combined with the 64 MeerKAT dishes. For SKA1-SUR the
sensitivity of 60 SKA dishes combined with 36 ASKAP dishes is used.
For SKA1-MID the $1\sigma$ sensitivity is given as 63 $\mu$Jy
beam$^{-1}$ over 100 kHz for one hour of observing time. For SKA1-SUR
this value is 263 $\mu$Jy beam$^{-1}$ over 100 kHz after one hour. We
scale these sensitivities using the appropriate integration time and
velocity resolution. For the sensitivity scaling as a function of
angular resolution we use the noise scaling curve shown in the bottom
panel of Figure 1 of the SKA1 Imaging Science Performance document
\citep{braunska}.  These scalings take into account the effect of the
spatial tapering needed to achieve the desired resolution.  In
addition, we apply an overall efficiency factor of 0.9 as listed in
Appendix A of \citet{braunska}. At 1.4 GHz SKA1-MID will have a field
of view of 0.5 square degrees. SKA1-SUR will be able to map 18 square
degrees due to its wide-field phased array feeds.

The $5\sigma$ column density limits for SKA1-MID, SKA1-SUR and SKA2 are
given in Table~\ref{noise}. These values agree well with those listed
in \citet{popping2014} which were derived using simulations.  Note
that a column density limit is defined as the flux density limit times
the velocity width, so for a fixed observation time and angular
resolution, observing with a larger velocity channel width will lead
to a \emph{higher} column density limit. This is therefore distinctly
different from the case where an already detected signal is smoothed
to a larger channel width giving an improved sensitivity (but at the
cost of a coarser velocity resolution).  

For SKA2, we assume that it
has 10 times the natural sensitivity of SKA1-MID, and apply a 50\%
correction factor to take into account tapering and weighting. We
assume this factor is independent of angular resolution.

In addition to column densities, we also list in Table \ref{masses}
the limiting \HI\ masses that can be detected using the various
setups. Here we assume the sources are unresolved and have a top-hat
velocity profile with a peak flux equal to $5\sigma$ and a velocity
width of $50$ km s$^{-1}$. Optimal smoothing (i.e., a channel width
equal to the velocity width) is assumed. Mass limits for other
velocity widths can be deduced by simply scaling with the width of the
profile.

\begin{table}
\scriptsize
%\tiny 
\caption{$5\sigma$ \HI\ column density limits (cm$^{-2}$)\label{noise}}
\centering
\begin{tabular}{lllll}
\hline
beam & \multispan{2}{velocity resolution \hfil} & & \\
\cline{2-5}
size & 1 km s$^{-1}$ & 5 km s$^{-1}$ & 10 km s$^{-1}$ & 20 km s$^{-1}$\\
\hline
\hline
 & & \multispan{2}{\hfil \bf SKA1-MID \hfil}  & \\
\hline
$t=10^h$   & & &  & \\
\hline
$1''$  & 1.75E+21	& 3.91E+21 &	5.53E+21 &	7.83E+21\\
$3''$  & 1.63E+20	& 3.65E+20 &	5.16E+20 &	7.30E+20\\
$10''$ & 1.13E+19	& 2.52E+19 &	3.57E+19 &	5.05E+19\\
$30''$ & 1.31E+18	& 2.94E+18 &	4.16E+18 &	5.89E+18\\
\hline
$t=100^h$ &  &  &  & \\
\hline
$1''$  &	5.53E+20 &	1.23E+21 & 1.75E+21 &	2.47E+21\\
$3''$  &	5.16E+19 &	1.15E+20 & 1.63E+20 &	2.30E+20\\
$10''$ &	3.57E+18 &	7.99E+18 & 1.13E+19 &	1.59E+19\\
$30''$ &	4.16E+17 &	9.32E+17 & 1.31E+18 &	1.86E+18\\
\hline
$t=1000^h$ &  &  &  & \\
\hline
$1''$  &	1.75E+20 &	3.91E+20 &	5.53E+20 &	7.83E+20\\
$3''$  &	1.63E+19 &	3.65E+19 &	5.16E+19 &	7.30E+19\\
$10''$  &	1.13E+18 &	2.52E+18 &	3.57E+18 &	5.05E+18\\
$30''$ &	1.31E+17 &	2.94E+17 &	4.16E+17 &	5.89E+17\\
\hline
\hline
 & & \multispan{2}{\hfil \bf SKA1-SUR \hfil}  & \\
\hline
$t=10^h$ & & & &\\
\hline
$1''$ &	4.71E+21&	1.05E+22&	1.49E+22&	2.11E+22\\
$3''$ &	4.45E+20&	9.96E+20&	1.40E+21&	1.99E+21\\
$10''$&	4.48E+19&	1.00E+20&	1.41E+20&	2.00E+20\\
$30''$&	8.38E+18&	1.87E+19&	2.65E+19&	3.75E+19\\
\hline
$t=100^h$ &  &  &  & \\
\hline
$1''$ &	1.49E+21&	3.33E+21&	4.71E+21&	6.67E+21\\
$3''$ &	1.40E+20&	3.15E+20&	4.45E+20&	6.30E+20\\
$10''$&	1.41E+19&	3.16E+19&	4.48E+19&	6.33E+19\\
$30''$&	2.65E+18&	5.93E+18&	8.38E+18&	1.18E+19\\
\hline
$t=1000^h$ &  &  &  & \\
\hline
$1''$ &	4.71E+20&	1.05E+21&	1.49E+21&	2.11E+21\\
$3''$ &	4.45E+19&	9.96E+19&	1.40E+20&	1.99E+20\\
$10''$&	4.48E+18&	1.00E+19&	1.41E+19&	2.00E+19\\
$30''$& 8.38E+17&	1.87E+18&	2.65E+18&	3.75E+18\\
\hline
\hline
 & & \multispan{2}{\hfil \bf SKA2\hfil}  & \\
\hline
$t=10^h$ & & & &\\
\hline
$1''$  & 1.13E+20&	2.52E+20&	3.57E+20&	5.05E+20\\
$3''$  & 1.25E+19&	2.80E+19&	3.97E+19&	5.61E+19\\
$10''$ & 1.13E+18&	2.52E+18&	3.57E+18&	5.05E+18\\
$30''$ & 1.25E+17&	2.80E+17&	3.97E+17&	5.61E+17\\	
\hline
$t=100^h$ & & & &\\
\hline
$1''$  & 3.57E+19&	7.99E+19&	1.13E+20&	1.59E+20\\
$3''$  & 3.97E+18&	8.87E+18&	1.25E+19&	1.77E+19\\
$10''$ & 3.57E+17&	7.99E+17&	1.13E+18&	1.59E+18\\
$30''$ & 3.97E+16&	8.87E+16&	1.25E+17&	1.77E+17\\
\hline
$t=1000^h$ & & & & \\
\hline
$1''$  & 1.13E+19&	2.52E+19&	3.57E+19&	5.05E+19\\
$3''$  & 1.25E+18&	2.80E+18&	3.97E+18&	5.61E+18\\
$10''$ & 1.13E+17&	2.52E+17&	3.57E+17&	5.05E+17\\
$30''$ & 1.25E+16&	2.80E+16&	3.97E+16&	5.61E+16\\
\hline
\hline
\end{tabular}
\end{table}

\begin{table}
\small
\caption{\HI\ mass limits for unresolved sources\label{masses}}
\vspace{2pt}
\centering
\begin{tabular}{llllllll}
\hline
beam	& \multispan{3}{\hfil $M_{\rm HI}/D^2_{\rm Mpc}$ for SKA1-MID\hfil} && \multispan{3}{\hfil $M_{\rm HI}/D^2_{\rm Mpc}$ for SKA1-SUR\hfil}\\
 \cline{2-4}  \cline{6-8}
& $t=10^h$ &  $t=100^h$ & $t=1000^h$ && $t=10^h$ &  $t=100^h$ & $t=1000^h$ \\
\hline
1$''$  &2.6E+03 &8.3E+02 &2.6E+02 & &7.1E+03&		2.2E+03&		7.1E+02\\
3$''$  &2.1E+03	&7.0E+02 &2.2E+02 & &6.0E+03&		1.9E+03&		6.0E+02\\
10$''$ &1.7E+03	&5.3E+02 &1.7E+02 & &6.7E+03&		2.1E+03&		6.7E+02\\
30$''$ &1.8E+03	&5.6E+02 &1.8E+02 & &1.1E+04&		3.6E+03&		1.1E+03\\
\hline
\end{tabular}
\vspace{2pt}\\
\HI\ masses were calculated assuming a $5\sigma$ flux density limit
and \\ a velocity width of 50 km s$^{-1}$. Optimal smoothing is assumed.
\end{table}

The science described here focuses on \HI\ observations
of low-redshift galaxies ($z<0.1$) so only frequency band 2 (950-1760
MHz on SKA1-MID and 650-1670 MHz on SKA1-SUR) is needed.

\section{Gas, Star Formation and Dark Matter at High Resolution}

\subsection{Gas and Stars}

The transformation of gas into stars is one of the most important
processes in galaxy evolution. Understanding the conditions that
determine the efficiency of this process, and the associated physics,
is the goal of many observational and theoretical studies. They also
form important input into numerical models of galaxy formation and
evolution. This requires knowledge of these processes over a large
range in scales: from galaxy-sized scales where gas is transported
from the disk of the galaxy into the halo and back, to kpc-sized
scales where gas clouds are collapsing, via sub-kpc scales where
neutral gas cools and turns molecular in Giant Molecular Clouds (GMC),
to parsec scales where individual stars are formed.  The latter, very
small, scales can be directly observed in our Galaxy, while processes
happening at galaxy scales can be studied in external galaxies. Tying
together the processes happening at these two extreme scales is a
major challenge: in our Galaxy we lack the overview, while in external
galaxies we rarely have the required resolution and sensitivity to study these
processes in detail.

Over the last decade the number of high-resolution, multi-wavelength
studies of galaxies in the nearby universe has increased
dramatically.  The \HI\ observations were obtained as part of dedicated surveys such as 
THINGS \citep{walter08},
Little THINGS \citep{hunter},
VLA-ANGST \citep{ott},
FIGGS \citep{begum},
WHISP \citep{whisp},
HALOGAS \citep{heald11},
SHIELD \citep{shield}, and
LVHIS \citep{koribalski}.
These surveys have made possible new studies of the
conditions for star formation on kpc or even sub-kpc scales in a
larger number of galaxies (see the chapter by \citealt{blyth2014} for
a more extensive description of recent \HI\ surveys).  These have made
it possible to make the first steps towards bridging the gap between
the observations at the scales of stars and those at scales of
galaxies. With the SKA the next step can
be taken.

\citet{kenn89} studied the gas and H$\alpha$ content of a number of
nearby galaxies and found a power-law relation between the total gas
surface density and the H$\alpha$-derived SFR.  Over the years,
similar studies (see e.g., \citealt{cal12, cal13}), using different
measures for the gas surface density, and different star formation
tracers (such as ultra-violet, H$\alpha$, infra-red or combinations of
them) found similar relations, but with a large spread in the
parameters. This may reflect actual variations in the physics, but a
large part of this spread is very likely also due to choice of sample,
analysis and star formation tracers.

%In \cite{kenn89, Kennicutt1998, Martin2001} evidence was found for a
%star formation threshold (see also \cite{skillman87}). That is, the
%star formation efficiency (as traced by H$\alpha$) decreases
%dramatically once the gas surface density drops below a certain
%threshold, but star formation does not shut off completely (e.g.,
%\cite{fer98}). \cite{kenn89} relates this star formation threshold
%density to the Toomre $Q$-parameter and the stability of the
%disk. \cite{schaye04} and \cite{taylor05} suggest that the star
%formation threshold is related to the formation of a cold phase in the
%neutral ISM when a sufficiently high surface density is reached. This
%cold phase suddenly makes efficient cooling possible leading to
%efficient formation of the cold gas needed to form stars.

Recent studies of the star-forming disks of nearby galaxies found a
tight linear relation between the molecular gas surface density and
the SFR surface density \citep{Leroy2008,Bigiel2008}.  This can be
interpreted as stars forming from the molecular ISM at a constant
efficiency.  Note though that these results do not say anything about
the necessary conditions for star formation on the scales of GMCs or
smaller. With the analysis performed at a resolution of 750 pc, any
relation must be interpreted within the context of ``counting
clouds'', i.e., one has to assume that the GMCs have uniform
properties with the observed molecular surface density determined by
the beam filling factor of these clouds.  Higher resolutions are
needed to go beyond this limitation.

\begin{figure}[tbp]
\centering % \begin{center}/\end{center} takes some additional vertical space
\includegraphics[width=.6\textwidth,angle=0]{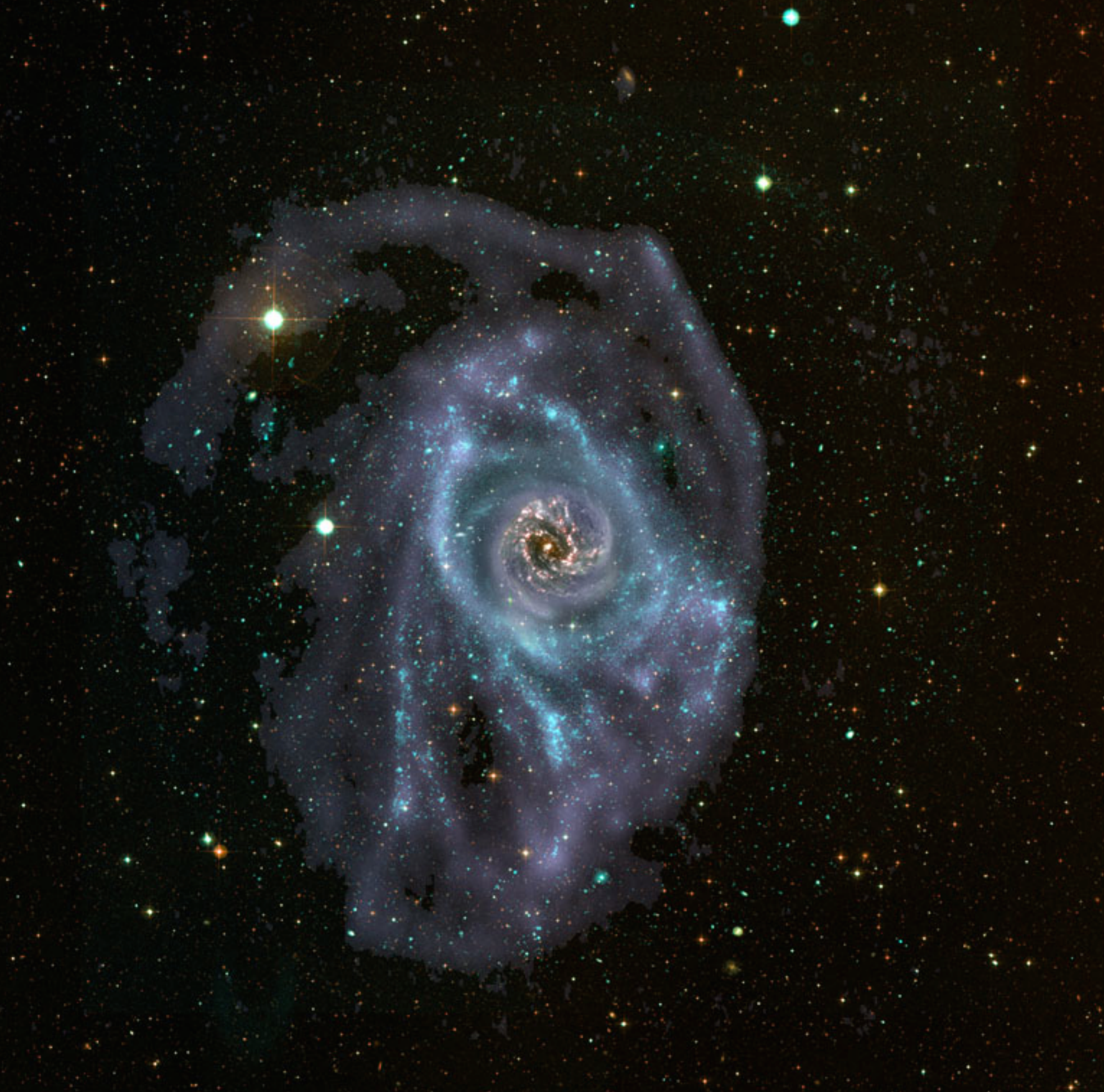}
\caption{\label{fig:m83} Composite multi-wavelength image of
  M83. Near- and far-UV from GALEX are shown in blue. Optical $R$-band
  from SDSS in green, $J$-band from 2MASS in red, \HI\ data from LVHIS
  is shown in pale-blue, and a combination of $B$, $R$ and H$\alpha$
  have been used as the luminosity of the stellar disk.  The field of
  view is $1.3^{\circ} \times 1.5^{\circ}$. (Picture courtesy
  B.\ Koribalski)}
\end{figure}

\citet{schruba11} found that the linear relation between molecular gas
and SFR surface densities extends into the \HI-dominated regime of the
outer disks of galaxies (see also \citealt{roychowdhury}).
%One interpretation of these results is that the Kennicutt-Schmidt law
%can be separated into two separate processes. The first one is star
%formation proceeding with a constant efficiency once H$_2$ is
%formed. The ``bottle neck'' determining the star formation rate thus
%seems to be the second process of conversion from HI into H$_2$
%(likely through the formation of a cold neutral component as
%discussed earlier).  The star formation threshold as found by
%\cite{kenn89} therefore seems to be caused by a change in SFR due to
%a changing ratio of HI to H$_2$ (as a function of total gas density),
%rather than an actual shutting off of star formation.
This is supported by the realisation that star formation in the outer
parts of disks is more widespread than originally thought.  GALEX UV
observations enabled direct detection of O and B stars which would
otherwise have escaped detection due to their inability to excite the
surrounding ISM enough to produce H$\alpha$ emission.  It is now
thought that these extended UV (XUV) disks are found in $\sim 30\%$ of
nearby disk galaxies \citep{thi07}. A striking example is M83 (NGC
5236) as shown in Fig.~\ref{fig:m83}. The outlying \HI\ structures
show up remarkably well in the UV, indicating that star formation is
progressing there as well, albeit with a lower efficiency
\citep{Bigiel2011} than in the bright, inner region of the galaxy.  An
extreme example of this is the dwarf galaxy ESO215-G?009
\citep{warren04}. This galaxy has the highest gas-to-light ratio known
for a galaxy in the low-redshift universe, yet, despite the large gas
reservoir, star formation apparently has halted, is inhibited or lacks
a trigger.

%A comparison of star formation as traced by H$\alpha$ with star
%formation as traced by UV radiation from O and B stars shows that
%while the level of H$\alpha$ emission typically drops very steeply at
%the edge of the optical disk (giving the appearance of a star
%formation threshold), such an edge is not observed in the UV
%emission. Rather, the density of young stars gradually decreases,
%again supporting the idea that the changing ratio of HI to H$_2$
%ultimately determines the overall star formation efficiency.

The above broad-brush picture illustrates the progress made in the
last few years, and mentions some of the empirical relations that are
now routinely used as input for numerical models that attempt to
explain the variations of star formation efficiency with, e.g.,
redshift, environment or galaxy mass.  It is however still difficult
to link the observations with the actual physical processes driving
the star formation rate.
For example, \citet{Leroy2008} test a number of theoretical explanations proposed
in the literature linking the gas density and the SFR. They looked at
the disk free fall time, the orbital timescale, the effects of
cloud-cloud collisions, the assumption of fixed GMC star formation
efficiency, and the relation between pressure in the ISM and the
phases of the ISM.  
%Similarly, they investigated several models
%proposed for the star formation threshold, such as gravitational
%stability in a purely gas disk, gravitational stability in a mixed
%disk of gas and stars, the effects of shear in a disk, and the onset
%of a cold gas phase.
Their conclusions were that none of these offers a unique explanation
for the observed behavior. So, though large-scale relations can be
identified (such as the Kennicutt-Schmidt Law), the actual physics is
a lot more complicated, and happens below the resolutions achieved so
far. Observationally, we will therefore have to probe the ISM at
scales below that of GMCs and individual \HI\ complexes. We also need
to have a better understanding of the balance between warm and cold
\HI\ phases, the efficiency of H$_2$ formation and the effects of
shocks and turbulence. Examples of the importance of high resolution
and high sensitivity for the study of turbulence in \HI\ are found in,
e.g., \citet{dutta,dutta2013}.

\citet{younglo96, younglo97}, \citet{young2003}, \citet{dB6822},
\citet{ianja2012}, \citet{warren} and \citet{stilp12} all analyzed the
\HI\ emission velocity profiles of galaxies and found evidence for the
presence of cold and warm \HI\ components by decomposing the velocity
profiles into components with a high and a low velocity dispersion.
They found that the cold (low velocity dispersion) component is
usually located near star-forming regions, whereas the warm (high
velocity dispersion) component tends to be found along every line of
sight. This presumably tells us something about the conditions for the
warm \HI\ to cool and turn molecular, but the key limitation of these
studies was again spatial resolution. The improved capabilities of the
SKA will enable these studies to be repeated but now resolving the
individual gas complexes.  (Equivalent absorption line studies are
described in the chapters by \citealt{Morganti_ska} and \citealt{naomi}.)
%By fitting
%these profiles with Gaussians, they found evidence for a
%broad component with a dispersion ranging from about 8 to 13 km
%s$^{−1}$ as well as a much narrower component with a dispersion
%ranging from 3 to 5 km s$^{−1}$ and associated the narrow component
%with the CNM and the broad component with the WNM phases of the ISM. A
%similar study was made byfor the Local Group dwarf
%galaxy NGC 6822. They also found narrow and broad HI components with
%mean velocity dispersions of 4 km s$^{-1}$ and 8 km s$^{−1}$,
%respectively. 
%\cite{ianja2012} and \cite{warren}
%used stacking of velocity profiles get high signal to noise
%determinations of the fractions of cold and warm HI. 
This should thus provide observations of a large number of galaxies at
sufficient resolution to gauge the ability of and the conditions for
the ISM to form GMCs over a wide range of galaxy conditions and
environments.

\begin{figure}[tbp]
\centering % \begin{center}/\end{center} takes some additional vertical space
\includegraphics[width=.6\textwidth,angle=0]{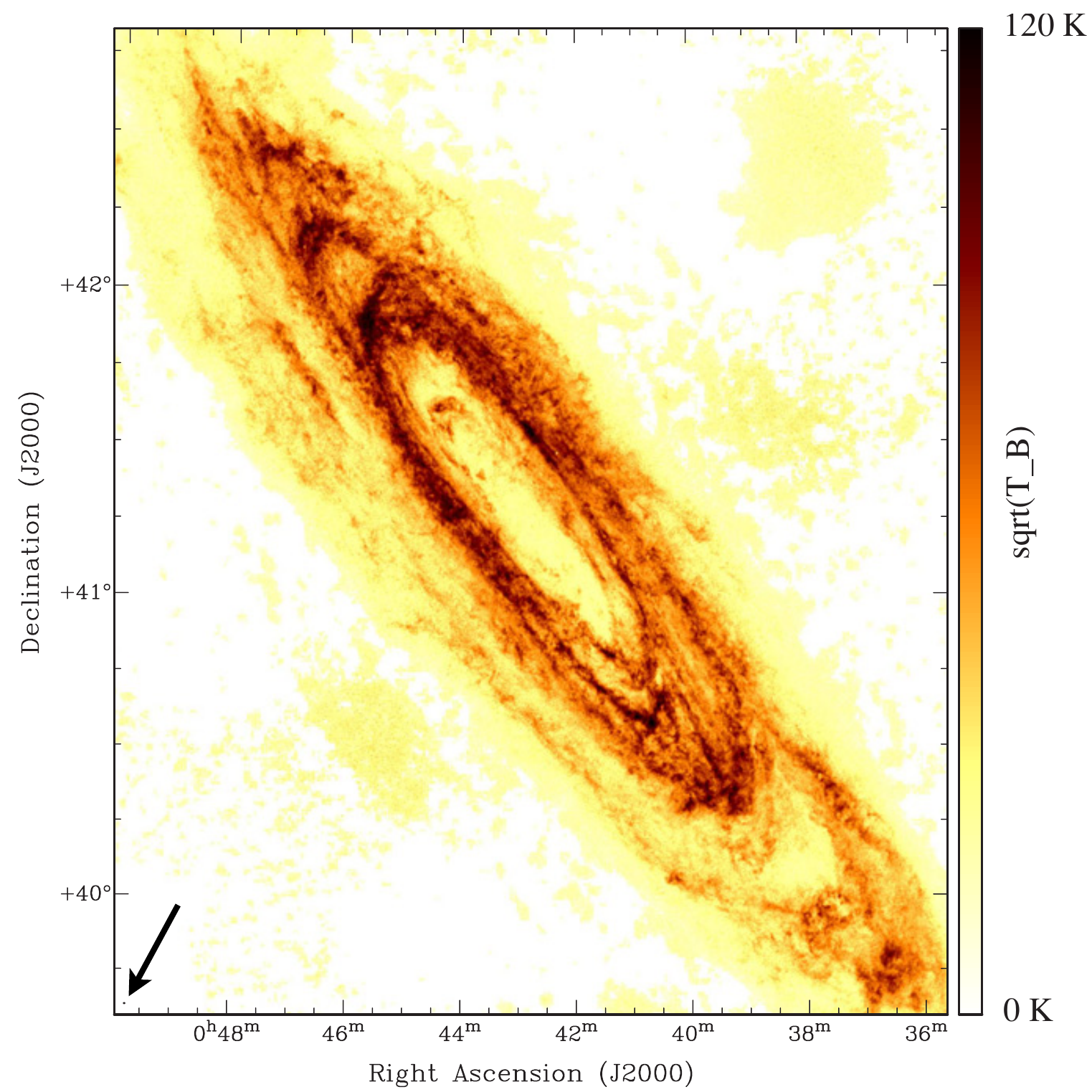}
\caption{\label{fig:m31} Integrated \HI\ map of M31 from
  high-resolution observations by \citet{braun2009}. The
  resolution of these observations is $\sim 50$ pc. The beam is
  indicated by the arrow in the bottom-left corner. SKA2 will enable
  highly detailed observations of this kind for hundreds of galaxies.}
\end{figure}

An angular resolution of $1''$ allows 100 pc
physical resolutions out to 20 Mpc (i.e., the distance of the Virgo
cluster). The high-resolution observations of M31 by \citet{braun2009} (shown in Fig.\ \ref{fig:m31})
have a maximum resolution of 50 pc, while the LMC observations by
\citet{kim2003} probe scales of 15 pc. Assuming SKA can observe
efficiently up to $\delta \sim +35^{\circ}$, then gives access to
$\sim 800$ galaxies within 20 Mpc that have independent distance
measurements. Of these $\sim 80\%$ have \HI\ detections. With the SKA
these can therefore all be studied at a resolution close to that of
the \citet{braun2009} M31 observations.  Out to 3 Mpc, we can study
galaxies at the same resolution of the \citet{kim2003} LMC
observations. There are $\sim 75$ galaxies with independent distance
measurements within that distance in the part of the sky accessible to
the SKA. This includes, for example, the Sculptor group. 

There are therefore a significant number of galaxies out to these
distances, and this opens up the exciting prospect of being able to
characterize the properties and morphologies of individual \HI\ clouds
in other galaxies over a wide range of environments. Combination with
high-resolution ALMA observations should provide a comprehensive
picture of the condition for and first phases of star formation.

\subsection{Dynamics and Dark Matter}

The high angular resolution that can be achieved with SKA will also
be of importance to studies of the internal dynamics of galaxies and
the distribution of dark matter.

\HI\ studies played a major role in the 1970s and 1980s to establish
the presence of dark matter in gas rich galaxies, necessary to keep up
the extended, flat (or rising) rotation curves at large radii. 
%It became readily clear, however, that additional dynamical methods
%are needed to establish accurately the amount of dark matter, due to
%the well known degeneracy between models with different stellar
%mass-to-light ratios of the visible parts. Even though some work
%argues for nearly maximum disks in Milky Way type galaxies, based on
%considerations of spiral structure or the strength of shocks in a few
%barred spirals, other work, primarily using stellar velocity
%dispersions, indicates non-maximum disks. This problem will continue
%to be addressed in the near future, with extensive surveys using
%integral field unit spectrographs (IFUs) on dedicated optical
%telescopes.
More recent work on late-type low surface brightness galaxies showed
that pure $\Lambda$CDM models are in disagreement with observations,
the so-called core-cusp controversy
\citep{2010AdAst2010E...5D}. Recent attempts to solve this problem
concentrate on introducing  feedback in the star formation
recipes used in the numerical simulations of galaxy formation
\citep{2011AJ....142...24O}, so that an initially cuspy dark matter
distribution can be modified sufficiently. Research on this problem
can be expected to continue to progress, until a better understanding
of these feedback mechanisms has been reached. At present, the
predictions from these simulations seem to begin to work for small
galaxies, but for larger ones the answer is still open. This is an
active area of research, and further improvements of the galaxy
formation models can be expected (e.g.,
\citealt{2014arXiv1404.5959D}).

In terms of further \HI\ observations addressing this problem, surveys
planned with the SKA precursors will enable the selection of an
adequate sample of relatively unperturbed galaxies, for which far
deeper \HI\ observations can be done with the SKA, as well as
molecular gas observations with ALMA.  These data will yield detailed
information on the circular and (equally important) non-circular
motions in these galaxies.  Combined with more extensive diagnostics
of the stellar kinematics using optical integral-field units, this
will yield crucial information about the galaxy kinematics, and hence
can be used to study the dark matter problem.  Moreover, they can also
be used for the purposes of studying the star formation and gas
accretion in disk galaxies, as discussed in this chapter.

%Another field whose exploration will florish in the near future is 
%the study of faint stellar extensions around the primary optical 
%image of galaxies, such as extended UV disks \citep{Espada11} or 
%faint optical emission (e.g., \citealt{Martinez-Delgado10}). The 
%spatial coincidence of these features with the HI is not always 
%unique (as in e.g. M33, \citealt{Lewis13}), and deep high resolution 
%imaging with the SKA will contribute to a better understanding of this 
%phenomenon, which may be related to interactions with and/or
%minor mergers of dwarf satellites. 

%In summary, the SKA will be able to probe nearby galaxies at
%significantly higher resolutions and sensitivities than was hitherto
%possible and will revolutionize our knowledge of the ISM in other galaxies.

\subsection{SKA Prospects\label{sec:sec2_3}}

Here we use the column density limits listed in Table \ref{noise} and
explore how the increased resolution and sensitivity of SKA makes new
science possible.  We use the results from the THINGS survey
\citep{walter08} as a benchmark. THINGS observed 34 nearby disk
galaxies with the combined VLA B, C and D array with a maximum angular
resolution of $6''$, or $\sim 500$ pc on average. For each galaxy,
observing time was 7$^h$ with the VLA B-array, 2.5$^h$ with
the C-array and 1.5$^h$ with the D-array. A total of $\sim 10^h$ is
therefore needed in this setup to produce a complete observation ---
this motivates our choice for $10^h$ as a typical observing time in
Sect.\ \ref{obssec}. With these parameters, the column density
sensitivity of the THINGS observations is $2.7 \cdot 10^{20}$
cm$^{-2}$ for a 5$\sigma$ detection over a 5 km s$^{-1}$ channel at
$6''$ resolution \citep{walter08}. This angular resolution is
effectively also the highest that can be achieved with the VLA B-array
at 1.4 GHz, and is currently the highest resolution at which \HI\ can
still be observed routinely.

To compare prospective SKA observations directly with THINGS, we focus
here on limits derived with a channel width of 5 km s$^{-1}$. A $10^h$
integration time at $6''$ with SKA1-MID will achieve a column density
limit of $8.1 \cdot 10^{19}$ cm$^{-2}$. This is close to the $4 \cdot
10^{19}$ cm$^{-2}$ THINGS sensitivity at $30''$.  A $10^h$ observation
on SKA1-MID at $6''$ thus enables studies at the resolution of THINGS,
but to column density limits that are a factor of 3 deeper, thus
probing the \HI\ well outside the star forming disks.  The larger
sensitivity of SKA1-MID allows tapering to spatial resolutions higher
than used by THINGS. For $10^h$ and at $3''$, the sensitivity at is
$3.7 \cdot 10^{20}$ cm$^{-2}$. This is comparable to the THINGS
sensitivity at $6''$ and therefore allows study of the star-forming
disk at twice the spatial resolution.  For galaxies out to 10 Mpc
the linear resolution corresponding to $3''$ is better than $\sim 145$
pc, which is starting to resolve individual \HI\ complexes.

With the current baseline design, $3''$ is an upper limit in terms of
resolution on SKA1-MID.  There is not enough sensitivity on the longer
baselines to push the resolution higher.  For example, performing the
same $10^h$ observation described above at $1''$ gives a column
density limit of $3.9 \cdot 10^{21}$ cm$^{-2}$, which is higher than
the maximum face-on column densities typically found in galaxies at
scales of hundreds of parsecs ($\sim 10^{21}$ cm$^{-2}$;
\citealt{Bigiel2008}). It is possible that optical depth effects mask
higher column densities at much smaller scales (e.g.,
\citealt{braun2009}), but this has been tested only in a limited
number of galaxies, and is the kind of project one would tackle with
the SKA for a larger sample.

Using SKA1-MID to make a longer observation of $100^h$, we reach
column density limits of $1.2 \cdot 10^{20}$ cm$^{-2}$ at $3''$
resolution (again using a channel spacing of 5 km s$^{-1}$).  This is
a factor of $\sim 2-3$ deeper than THINGS at twice the spatial
resolution, and will enable high-resolution characterization of the
ISM at scales better than 145 pc out to 10 Mpc.

Even with an increased observing time, resolutions of $1''$ remain
challenging.  At $100^h$ the limit is $1.2 \cdot 10^{21}$ cm$^{-2}$
which is comparable to the highest column densities typically found in
galaxies. At 1000 hours, the corresponding limit is around $3.9 \cdot
10^{20}$ cm$^{-2}$, and this would allow mapping of most of the
\HI\ disk. This seems, however, a large investment of time for a single
galaxy.  For high-resolution observations of \HI\ in nearby galaxies
using SKA1-MID, $3''$ thus seems to be the practical (design-imposed)
limit.

For a 50\% SKA1-MID, many aspects of the science mentioned above can
still be done. The noise will be a factor two higher, but this still
allows $3''$ resolution observations that, even at $10^h$, lead to
sensitivities that are better than THINGS at $6''$.

At $10^h$ observing time, the $5\sigma$ limits for SKA1-SUR at
resolutions better than $5''$ are too high for mapping galaxies. For
resolutions between $\sim 10''$ and $\sim 30''$ the sensitivities are
high enough to perform THINGS-like observations.  The strength of the
SKA1-SUR observations will, however, be in their ability to
characterize the environment due to its large instantaneous
field-of-view of 18 square degrees.

The increased sensitivity of a SKA2 means that an order of
magnitude increase in column density sensitivity can be achieved with
respect to SKA1-MID.  A $10^h$ observation will reach down to $2.8
\cdot 10^{19}$ cm$^{-2}$ at $3''$ and 5 km s$^{-1}$ resolution. At
$1''$ and 5 km s$^{-1}$ resolution, the column density sensitivity
that is achieved is comparable to that of THINGS (at $6''$), but with
a spatial resolution that is a factor of 6 better.

At $100^h$, $3''$ observations with SKA2 reach levels around $8.9 \cdot
10^{18}$ cm$^{-2}$, with even a $1''$ beam giving limits of $8.0 \cdot
10^{19}$ cm$^{-2}$, i.e., enough sensitivity to reach the levels
typically found in the outer disks of galaxies. The \citet{braun2009}
observations of M31 (Fig.\ \ref{fig:m31}) have a linear resolution of
50 pc. With SKA2 at $1''$ we can thus reach similar resolutions out to
10 Mpc. This opens the exciting prospect of mapping galaxies at a
physical resolution comparable to the M31 observations shown in
Fig.\ \ref{fig:m31}. Furthermore, the SKA2 would make it possible to do
this for hundreds of galaxies. Characterisation of the ISM at a
resolution of tens of parsecs in such a large number of galaxies, each
presumably with different physical conditions, would dramatically
improve our knowledge of the ISM in other galaxies.

\section{The Galactic Fountain and Accretion}

\subsection{Background}

Galaxies are not isolated systems, as they continuously interact and
exchange material with the environment in which they reside.
Understanding how and in what form gas is transferred to galaxies from
the IGM and vice-versa is a major challenge for current cosmological
models.  Star-forming galaxies are expected to accrete fresh gas to
feed their star formation throughout the whole Hubble time.  Evidence
for gas accretion includes estimates of the gas depletion times
\citep{Bigiel2011}, the study of the star formation histories
\citep{Panter+2007} and chemical evolution models of the Milky Way
\citep{Chiappini+1997, Schoenrich&2009}.  Indirect determinations of
the gas accretion rate show that it should closely follow the SFR in
every galaxy \citep{Hopkins+2008, Fraternali&2012}.  However, there is
little evidence for the majority of the gas accretion taking place in
the form of gas clouds infalling into local galaxies.  In the Milky
Way, the accretion rate from high velocity clouds is only 0.08
$M_{\odot}\,{\rm yr}^{-1}$ \citep{putman}, more than an order of
magnitude lower than the SFR ($1-3\ M_{\odot}\,{\rm yr}^{-1}$,
\citealt{Chomiuk&2011}).  \HI\ studies of nearby galaxies give similar
discrepancies \citep{sancisi}.

\begin{figure}[tbp]
\centering % \begin{center}/\end{center} takes some additional vertical space
\includegraphics[width=.95\textwidth,angle=0]{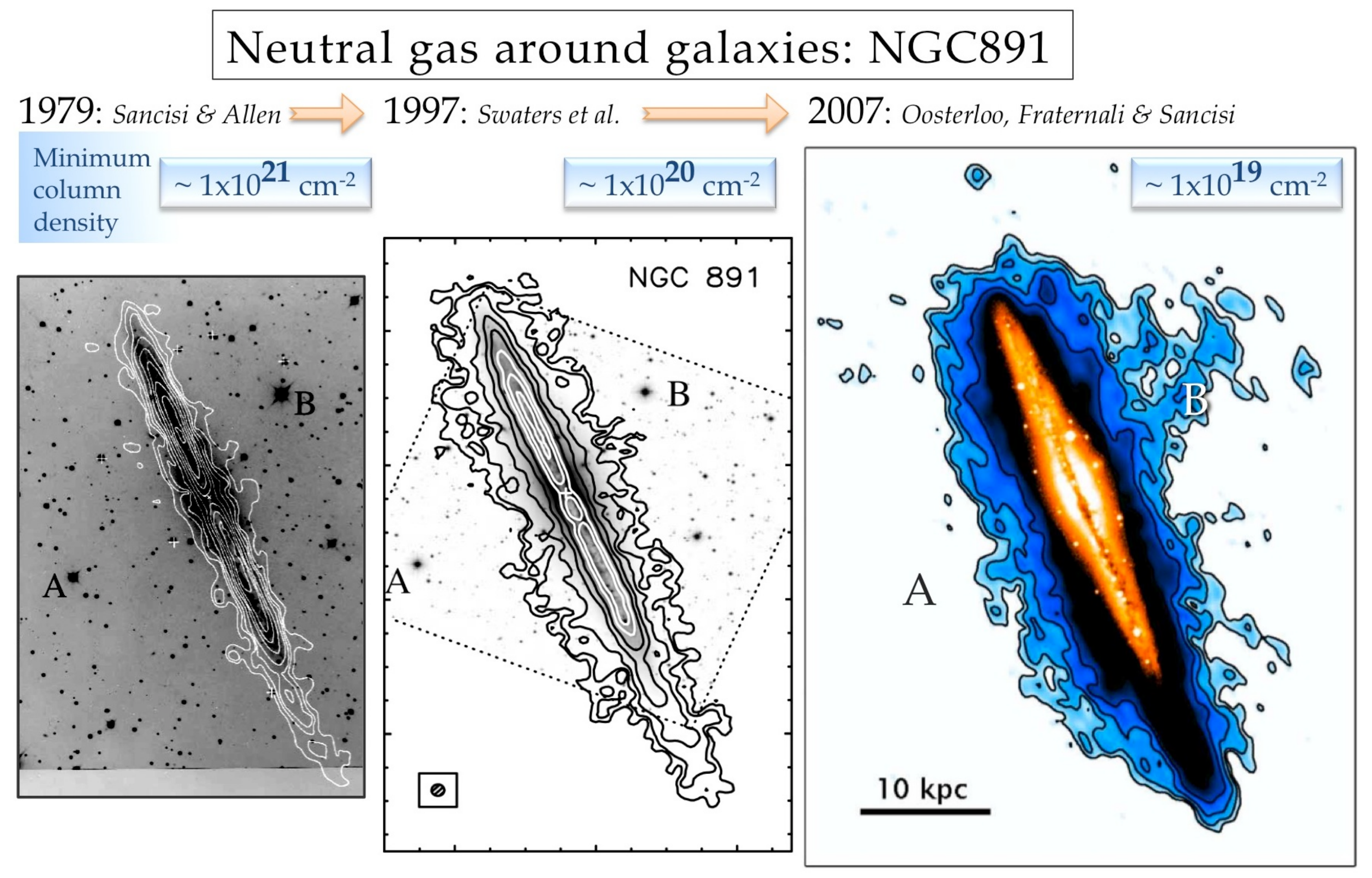}
\caption{\label{fig:n891progress} Three generations of neutral
  hydrogen observations of the edge-on spiral galaxy NGC\,891,
  obtained with the Westerbork Synthesis Radio Telescope in 1979
  \citep{Sancisi&1979} (left panel), 1997 \citep{Swaters+1997} (middle
  panel) and 2007 \citep{oosterloo} (right panel). The improvement in sensitivity of
  only one order of magnitude between each observation led 
  to the discovery of the \HI\ halo (middle panel) and next to the
  realization that this halo comprises almost 30\% of the \HI\ mass of
  the gas and hosts gas filaments extending up to more than 20 kpc
  from the disk (right panel). SKA will go down two more orders of
  magnitude in column density, we can only wonder what we will find.}
\end{figure}

A possibility is that accretion takes place at very low column
densities.  At the moment the deepest \HI\ observations of nearby disk
galaxies obtained with an interferometer reach column densities
slightly below $10^{19}$ cm$^{-2}$ \citep{oosterloo, walter08,
  westmeier11, heald11}.  Fig.\ \ref{fig:n891progress} shows three
observations of the same galaxy, NGC\,891, obtained in the last thirty
years with the Westerbork Synthesis Radio Telescope (WSRT).  The
improvement in sensitivity from $\sim 1 \cdot 10^{21}$ cm$^{-2}$ (left
panel; \citealt{Sancisi&1979}) to $\sim 1 \cdot 10^{19}$ cm$^{-2}$
(right panel; \citealt{oosterloo}) has revealed a completely different
picture of the \HI\ distribution in this galaxy.  The most recent data
have shown the presence of an extended \HI\ halo which everywhere
extends up to $8-10$ kpc above the disk of NGC\,891, and with a long
filament extending up to 20 kpc.  Most of this gas is likely to be
produced by stellar feedback from the disk
\citep{Fraternali&Binney2008} but a fraction could come from
accretion. The long filament is, however, likely due to a minor
interaction (cf.\ \citealt{oosterloo, mapelli}).  Clouds and filaments
at similar column densities are observed also around other galaxies
\citep{Fraternali+2002,mundel,westmeier05,sancisi}.  As an aside, a
similar increase in the prominence of \HI\ when going to lower column
densities is also found for early-type galaxies. Here the \HI\ detection
rate outside clusters increases from $\sim 0\%$ to $\sim 40\%$ when
going from $1 \cdot 10^{21}$ cm$^{-2}$ to $1 \cdot 10^{19}$ cm$^{-2}$
\citep{paolo}.

Gas at lower column
densities is therefore certainly present around galaxies.  This is
indicated by deep single-dish observations (e.g.,
\citealt{deblok2014}), and by the study of Lyman-$\alpha$ absorbers
towards quasars and their association with the circumgalactic media of
nearby galaxies \citep{Tumlison+2013}. The distribution of this
ultra-low column density gas can only be directly measured by the SKA.

%Competing models are now proposed to solve the long-standing problem
%of sustainability of star formation in disks.  Cosmology predicts that
%large nearby galaxies should be surrounded by hot coronae that contain
%a significant fraction of the so-called missing baryons
%\cite{Bregman2007,Crain+2010}.  Hydrodynamical simulations suggest
%that these coronae should be, under certain conditions, thermally
%unstable and cool to form cold clouds \cite{Kaufmann+2009} but this
%idea has been challenged by other theoretical work \cite{Binney+2009}.

The presence of low-column density gas around galaxies is also
predicted by models. For example, some recent cosmological simulations
show {\it cold} gas filaments penetrating the hot halos of galaxies
and reaching the disks to feed star formation (e.g.,
\citealt{Fernandez+2012}), though the magnitude of the implied
accretion is still a matter of some debate \citep{Nelson+2013}.
Finally, galactic fountain models predict that the hot gaseous halos
of galaxies are polluted and cooled by stellar feedback from the
galactic disks \citep{Fraternali&Binney2008, Marinacci+2010}.  All
these models make predictions on the presence and distribution of low
column density material that can be directly tested by SKA (see also
the chapter by \citealt{popping2014}).

SKA will as well reveal the fate of the ``missing'' \HI\ in more
high-density environments such as Hickson Compact Groups
\citep{verdes, borthakur}, which is suspected to be in the form of
low-column density intra-group material extending to large relative
velocities.

Given the dramatic difference which two orders of magnitude make, as
illustrated in Fig.\ \ref{fig:n891progress}, it is difficult to
anticipate what the SKA2 will be able to see when going down to a
sensitivity of below $10^{17}$ cm$^{-2}$.  However, what is clear, is
that with SKA we will obtain the full picture of \HI\ around galaxies,
characterize the properties and the kinematics of the \HI\ halos, and
derive precise estimates of gas accretion. SKA will provide a
definitive answer on how gas falls into galaxies from intergalactic
space.

\subsection{Current Status}

\begin{figure}
\centering
\includegraphics[width=0.32\textwidth]{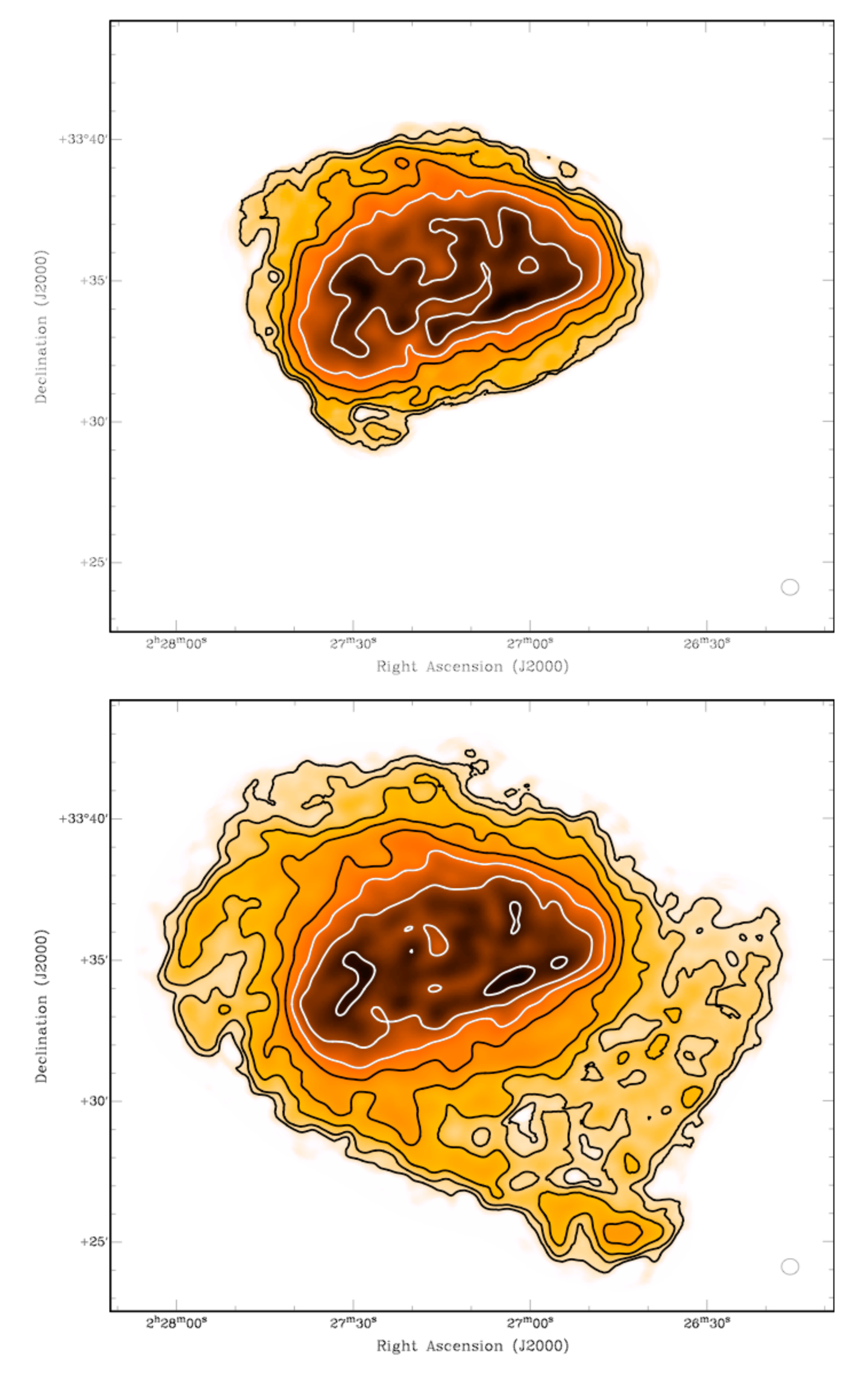}\includegraphics[width=0.68\textwidth]{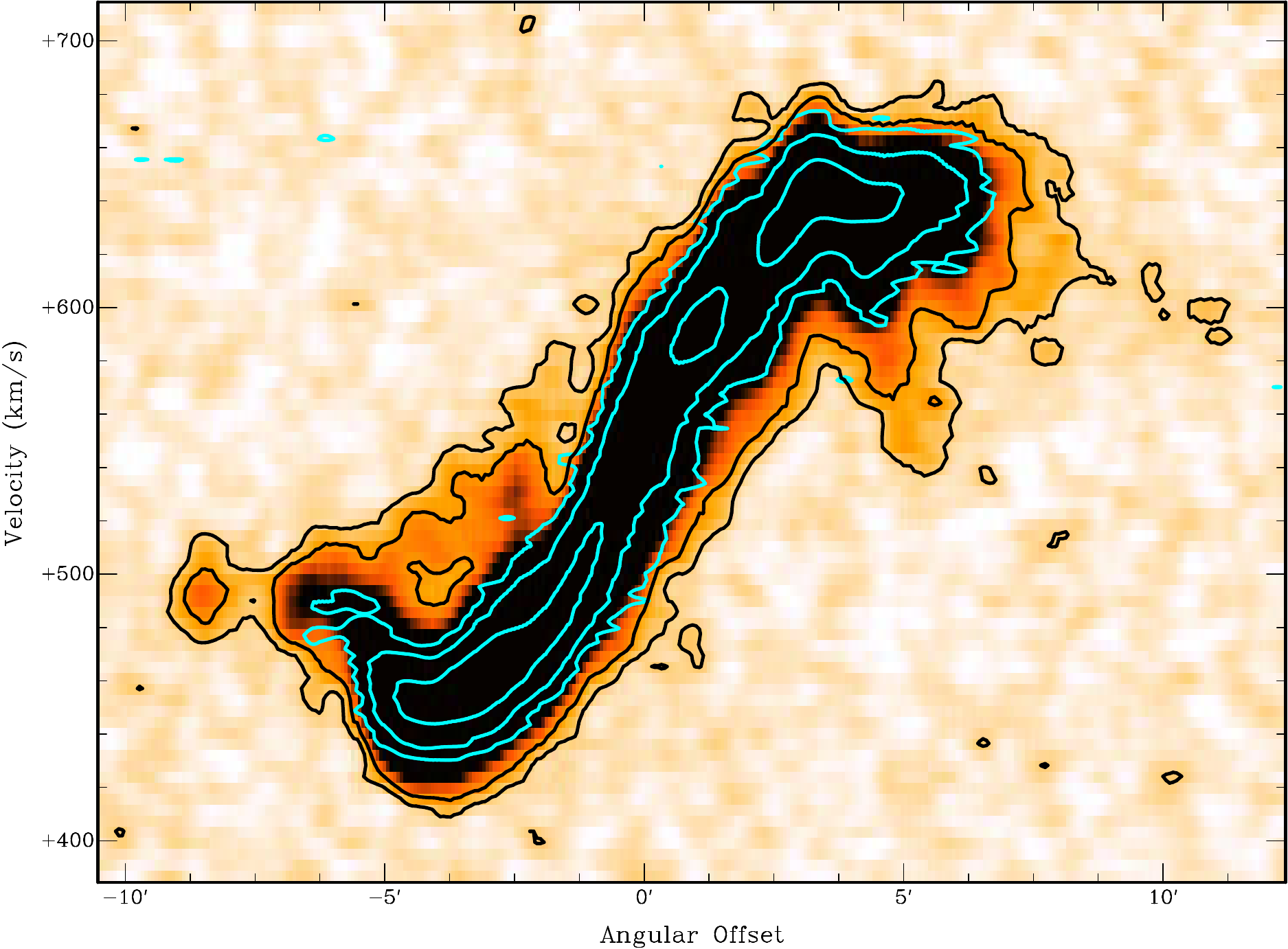}
\caption{Comparison between THINGS and HALOGAS data for NGC 925. The
  THINGS data have been smoothed to the same resolution
  ($\sim30^{\prime\prime}$) as the HALOGAS data. Top left: THINGS
  moment-0. The lowest contour is at $5\cdot 10^{19}$ cm$^{-2}$, and
  contour levels increase by factors of two.  Bottom left: HALOGAS
  moment-0; the lowest contour is $1 \cdot 10^{19}$ and levels
  increase by factors of two.  Right: PV diagram near the major axis
  (grayscale and black contours from HALOGAS at $3,9\sigma$, cyan from
  THINGS at $3,9,27\sigma$). This figure shows how the increased
  column density sensitivity of HALOGAS enables mapping of the low
  column density gas in and around galaxies.}
\label{figure:comparison}
\end{figure}

The WSRT Hydrogen Accretion in Local Galaxies (HALOGAS) survey
\citep{heald11} aims to catalogue accreting gas in nearby spiral
galaxies, and to determine the ubiquity, properties, and origin of
neutral gas halos such as the one associated with NGC~891
(Fig.\ \ref{fig:n891progress}). The HALOGAS survey observed 24
galaxies for 120 hours each to a column density limit of $\sim 1 \cdot
10^{19}$ cm$^{-2}$ (assuming a $5 \sigma$ limit and a linewidth of 12
km s$^{-1}$). As noted above, this level is suitable for confidently
detecting and characterizing diffuse neutral extraplanar gas, and also
for identifying isolated clouds around the target galaxies. The
HALOGAS detection limit for unresolved cloud masses is approximately
$M_\mathrm{HI} \approx 2.7 \cdot
10^5\,(D/10\,\mathrm{Mpc})^2\,M_\odot$. The HALOGAS observations with
the WSRT attain the best compromise between angular resolution and
sensitivity at around $30^{\prime\prime}$, meaning that for a galaxy
at a typical distance of 10 Mpc, the linear resolution is around 1.5
kpc. While a higher resolution survey like THINGS is better suited for
study of small-scale structure in the ISM and the connection to star
formation (as discussed in the previous section), the HALOGAS
observations are much more sensitive to diffuse structures (see as an
example data for NGC 925 in Fig.\ \ref{figure:comparison}).

HALOGAS provides the capability for a uniquely deep and uniform search
for gas clouds and streams in the environment of the target
galaxies. These clouds and streams, if present, could represent the
most obvious signposts of ongoing cold accretion in the local
Universe. A complete accounting of the potentially accreting gas is
still being finalized \citep{juette_etal_inprep} but it can already be
concluded that the total visible accretion rate attributable to such
features is rather small. HALOGAS shows that a substantial population
of gas clouds and streams in neutral gas is not present down to the
$\sim 10^{19}$ cm$^{-2}$ survey column density limit.  There are clear
consequences for the mechanisms that can be invoked to explain the
continued fuelling of star formation and the column density levels at
which these occur.

%If cold accretion is not the main process by which
%galaxies acquire gas, then alternatives such as galactic fountain
%driven positive feedback \citep{Fraternali&Binney2008, Marinacci+2010},
%would be the relevant observational tracer of the gas accretion
%process for spirals in the local Universe. This result bears
%confirmation and further investigation with the SKA.

Crucially, the benefit of deep observations is not just the
sensitivity to faint diffuse outer disk emission, but also to faint
diffuse thick disk gas seen in projection against the inner thin (main) disk.
%For bright edge-on galaxies,
%the star formation intensity is such that a thick \HI\ disk is set up,
%whose kinematics shows rotation speeds which are smaller than those in
%the plane of the disk (e.g.,  \citealt{oosterloo}) for NGC 891. 
%This
%greatly complicates the analysis of flaring \HI\ disks, which are
%expected to exist in the outer parts of galaxies if the dark halo is
%near spherical, and which can help constrain the shape of the dark
%halo. Perhaps this problem can still be studied for a number of small
%galaxies, which have a more quiescent star formation activity.
%The unprecedented and uniform depth of the HALOGAS observations also
%allow some broad conclusions about the nature of the vertical
%structure in disk galaxies to be drawn for the entire sample. Some
%galaxies, like NGC 925 \cite{heald11} (as shown in Figure
%\ref{figure:comparison}) along with NGC 3198 \cite{gentile_etal_2013}
%and NGC 4414 \cite{deblok_etal_2014} are seen to host thick \HI\ disks
%like NGC 891 \cite{oosterloo}. On the other hand, some other
%sample galaxies like NGC 4244 \cite{zschaechner_etal_2011} do
%not. {\it These non-detections provide crucial leverage to disentangle
%  the origin of such structures.} 
Analysis of the HALOGAS sample shows that the key discriminating
factor in determining the presence of such a thick \HI\ disk is the star
formation rate normalized by the disk area
\citep{heald_etal_2014}. These thick disks in many cases show rotation
speeds which are smaller than those in the plane of the disk (e.g.,
\citealt{oosterloo}).
%This conclusion is similar to the analysis of
%the origin of radio continuum halos \cite{dahlem_etal_2006}, which
%seem to be also be present when sufficient star formation energy input
%is available per unit area to overcome the disk gravity (as traced by
%the stellar mass). 
The dependence on star formation rate surface density points strongly
toward the galactic fountain being the dominant factor in creating
thick \HI\ disks in galaxies. It looks consistent with the picture of
galactic-fountain-driven positive feedback
\citep{Fraternali&Binney2008, Marinacci+2010}. The presence and
importance of extra-planar material (such as a thick \HI\ disk) would
then be the relevant observational tracer of the gas accretion process
for galaxies in the local Universe. This result bears confirmation and
further investigation with the SKA.

\subsection{SKA Prospects}

The current configuration of SKA1-MID is heavily dominated by short
baselines and therefore has an exquisite column density sensitivity at
moderate resolutions. To compare its performance with existing
observations we use the HALOGAS survey \citep{heald11} as described
above as a benchmark.  HALOGAS reaches a typical $5\sigma$ column
density sensitivity of $5 \cdot 10^{18}$ cm$^{-2}$ per 4.1 km s$^{-1}$
channel at a spatial resolution of $30''$.  This beam size represents
the optimal combination between resolution and sensitivity for these
particular WSRT observations and is the standard resolution HALOGAS
used.

We compare these numbers with those derived for the SKA as shown in
Table \ref{noise}. For a $10^h$ observation with SKA1-MID using a
30$''$ beam and 5 km s$^{-1}$ velocity resolution, the resulting
$5\sigma$ column density limit is $2.9 \cdot 10^{18}$ cm$^{-2}$ . In
other words, a $10^h$ SKA1-MID observation already goes twice as deep
as HALOGAS. This therefore opens the prospect of routinely obtaining
deep observations of a large number of galaxies in a modest amount of
time, and efficiently characterizing the presence of clouds, thick
\HI\ disks and other extra-planar features. This will lead to a more
complete statistical description of the importance and mechanisms for
accretion.

With $100^h$ of observing on SKA1-MID (again at $30''$ and 5 km
s$^{-1}$), a column density limit of $9.3 \cdot 10^{17}$ cm$^{-2}$ is
reached. This is comparable to the column densities that deep
\HI\ observations of nearby galaxies with the Green Bank Telescope
reach \citep{pisano14}, but with a spatial resolutions that is almost
a factor 20 better (from $9'$ to $30''$).

A 50\% SKA1-MID will still be able to reach HALOGAS depth, but in
$10^h$ per galaxy instead of $120^h$.  In terms of increasing the
statistics of accretion processes these observations will therefore
already be scientifically valuable.

For a fixed observing setup, the column density sensitivity of
SKA1-SUR is lower by a factor of 2.7 (for $3''$) to 6.4 (for $30''$).
Though observations will thus be less sensitive, the large field of
view will help exploring the extended environment around nearby
galaxies, and can help placing the SKA1-MID results in the context of
the large-scale cosmic web.

For SKA2, at a resolution of $30''$, the column density limits
after $10^h$ are truly impressive, reaching $\sim 2.8 \cdot 10^{17}$
cm$^{-2}$, i.e., more than an order of magnitude deeper than HALOGAS.
At $100^h$, these numbers go down even further to $\sim 8.9 \cdot
10^{16}$ cm$^{-2}$. At these column densities one would be able to
detect the 21-cm emission from Lyman Limit Systems. Further discussion
on this and the cosmic web in general is presented in the chapter by
\citet{popping2014}.

\section{Accretion and external high-velocity clouds}

\subsection{HVCs as accretion probes}

The previous section emphasised the importance of the extra-planar gas
as a tracer of accretion. Here we discuss some of the properties of
the clouds that form part of that extra-planar component. In our Milky
Way (MW) these are observed as HVCs (see the chapter by \citealt{naomi}
elsewhere in this volume). Here we discuss the prospects of observing
these clouds in other galaxies.

The physical properties of the HVCs around the MW span a large range;
the large complexes are at distances of $\sim10$ kpc, with similar
heights above the Galactic plane, and have physical sizes of several
kpc and masses of $1-50\times10^5$ $M_{\odot}$
\citep{putman,2008ApJ...672..298W,2007ApJ...670L.113W}.  Compact
clouds near the disk-halo interface are $\lesssim 1$ kpc above the
disk and have sizes and masses of $\sim 700$ $M_{\odot}$ and $\sim 30$
pc \citep{2010ApJ...722..367F}.  In addition, there may be populations
of clouds distributed throughout the halo of the MW. These compact
high velocity clouds (CHVCs) appear to be circumgalactic and at
distances of $\sim50-150$ kpc with masses of $\sim 5 \times 10^4$
$M_{\odot}$ and sizes of $\sim 0.8$ kpc for an assumed distance of 100
kpc
\citep{2002ApJS..143..419S,2003ApJ...589..270M,2002AJ....123..873P,2002A&A...392..417D}.
The difference in observed properties of the HVCs is a clue to their
likely origins.  The velocities of the large complexes and their low
metallicities ($\sim 0.15\, Z_{\odot}$) are consistent with accretion
of gas onto the MW \citep{2007ApJ...670L.113W}.  The spatial
distribution and kinematics of the clouds at the disk-halo interface
are consistent with a galactic fountain \citep{2010ApJ...722..367F}.
The same phenomenon is traced by the IVCs. The
CHVCs potentially provide information on the Galactic dark matter
substructure \citep{2002ApJS..143..419S}.

HVCs are thus a probe of both accretion in galaxies and the recycling
of gas through the hot halo as part of the galactic
fountain. Observing and understanding these processes in our own
Galaxy is hampered by the difficulty of constraining the distance to
these systems. In external galaxies, we can directly determine the
masses and sizes of the HVCs and have the opportunity to view the HVCs
in relation to the galactic disks. Unfortunately, attempts to search
for HVCs in external galaxies are often hampered by lack of resolution
and sensitivity.  HVCs are low column density structures with peak
column densities of $\sim 10^{19}$ cm$^{-2}$.  The compact clouds
(disk-halo and CHVCs) will be point source detections for most
extragalactic surveys with masses $< 10^5$ $M_{\odot}$ down to $\sim
100$ $M_{\odot}$ for clouds at the disk-halo interface.

%In addition, in order to trace the HI associated with the galactic
%fountain and accretion one wishes to observe HVCs that are \emph{not}
%associated with the interaction of satellite galaxies.  If viewed
%externally, the dominant HVC structure of the MW that would be visible
%is the Magellanic Stream, which is associated with the infall and
%accretion of the Magellanic Clouds. Indeed, a majority of anomalous HI
%seen in external galaxies can typically be attributed to accretion of
%satellite galaxies or tidal interactions \cite{sancisi}.

Only a handful of nearby disk galaxies outside the Local Group have
detections of discrete clouds of anomalous \HI: NGC 891 with clouds of
$1-3 \times 10^6$ $M_{\odot}$ \citep{oosterloo}, NGC 2403 with clouds
of $6-10 \times 10^6$ $M_{\odot}$ \citep{2001ApJ...562L..47F}, NGC
2997 with clouds of $\sim 10^7$ $M_{\odot}$
\citep{2009ApJ...699...76H} and NGC 55 with clouds of $2-40 \times
10^6$ $M_{\odot}$ \citep{2013MNRAS.434.3511W}.  These anomalous
\HI\ clouds appear to be the analogs of the large HVC complexes in the
MW with substantial \HI\ masses, large extents of several kpcs and
projected separations from the galactic disks of $\sim 10-20$ kpc.
%The only exception is deep
%observations of M31 which revealed $\sim 20$ HI clouds with masses of
%$\sim 3 \times 10^5$ $M_{\odot}$; even then the peak column densities of
%these clouds are a factor of $\sim 4$ higher than the typical peak
%column densities of the MW HVCs \cite{2004ApJ...601L..39T,
%  westmeier05}.  Indeed, multiple 
Other dedicated surveys of other
nearby galaxies and galaxy groups reveal no HVC
analogs down to $\sim 4 \times 10^5$ $M_{\odot}$ at larger distances from galaxies
\citep{pisano07, 2009ApJ...692.1447I}. 

Further systematic studies of nearby galaxies are thus needed to
determine the properties of the global HVC population.  Observations
much deeper than those that currently exist are needed to extend the
extragalactic HVC population to the low mass clouds at the disk-halo
interface that are associated with the galactic fountain mechanism.

\subsection{Prospects for HVCs with the SKA}

Table \ref{tab:hvcs} summarizes the properties of HVCs.  The large HVC
complexes have physical sizes of $\sim 10$ kpc and will remain
resolved with a 30$''$ beam out to distances of $\sim 30$ Mpc, so
observing requirements in the nearby Universe will be set by column
density limits.  For typical $z$-heights of $\sim 10$ kpc, the HVC
complexes should be spatially resolved from the disk out to 10 Mpc for
all but the most face-on systems ($i>20^{\circ}$) at 30$''$
resolution.  In more face-on systems they can however be detected
because of their velocity separation from the main disk (e.g., in
M101; \citealt{kamphuis}). In a $10^h$ observation, SKA1-MID can
achieve a $5\sigma$ column density sensitivity of $\sim 5 \times
10^{18}$ cm$^{-2}$ in a 20 km s$^{-1}$ channel (matched to the typical
linewidth of the HVCs). This is a factor of 2 lower than the typical
peak column density for MW HVCs; at this level a survey of nearby
galaxies would be sensitive to the full range of MW HVC complexes. A
50\% SKA1-MID would be sensitive to the most massive systems with peak
column densities $\gtrsim 10^{19}$ atoms cm$^{-2}$.

SKA2 will offer greatly improved column density
sensitivity.  Increasing sensitivity to lower column density allows
the tracing of the extended structure of the HVCs which can offer
clues to the environment, including the pressure of the medium they
are embedded in. Higher resolution observations trace the clumpiness
and structure of the \HI\ complexes.

The compact clouds at the disk-halo interface are physically very
small and will be mostly unresolved point sources except at the
highest resolutions in the closest galaxies. With a typical size of 30
pc they have angular sizes of 6$''$ $D^{-1}_{Mpc}$. For a 25 km
s$^{-1}$ channel (matched to the linewidth of HVCs; Table \ref{masses}
scaled by velocity width), SKA1-MID with a 10$''$ beam is sensitive to
the typical disk-halo cloud of \citet{2010ApJ...722..367F} of $\sim
700$ $M_{\odot}$ at 1 Mpc in a 100$^h$ observation.  Deep observations
of very nearby galaxies ($D=2$ Mpc) would be sensitive to the most
massive disk-halo clouds ($M_{\rm HI} > 3000$ $M_{\odot}$).  With a
50\% SKA1-MID, the most massive disk-halo clouds would be detectable
out to 1.4 Mpc. SKA2 will be able to detect the median
disk-halo cloud population in deep (100$^h$) observations out to $\sim
3$ Mpc.

SKA1-SUR lacks the column density sensitivity to efficiently detect
the large HVC complexes in large surveys; neither does it have the
point source sensitivity to detect the disk-halo clouds. However, its
large field of view does make it ideal for deep surveys of nearby
galaxies to search for an extended circumgalactic HVC population
separated from the parent galaxy by multiple degrees.  With a 10 hr
observation at 30$''$ and 20 km s$^{-1}$, SKA1-SUR has a column
density sensitivity of $\sim 4 \cdot 10^{19}$ cm$^{-2}$. This would
allow for the detection of the high-$N_{\rm HI}$ tail of the CHVC
population.  At $100^h$ the sensitivity is $\sim 1.2 \cdot 10^{19}$
cm$^{-2}$; this is well-matched to the median peak $N_{\rm HI}$ of the
CHVC population.  With a 50\% SKA1-SUR, CHVCs with high peak column
densities could still be detected in deep 100$^h$ observations.

%Deep observations of a few nearby galaxies could help
%definitely constrain the distance (and hence masses and sizes) of the
%CHVC population.  CHVCs located close to the disk would be detected in
%the same observations described above for the large HVC complexes.

\begin{table}
\scriptsize
\caption{Galactic and Extragalactic HVCs \label{tab:hvcs}}
\begin{tabular}{lccccccl}
\hline
{Class} & {$d_{HI}$} & {$\theta_{HI}$} & {$N_{HI}$} & {$M_{HI}$} & {$z$-height} & {$\theta_z$} & References \\
{}& (kpc) & {($D_{Mpc}^{-1}$ kpc)} & (cm$^{-2}$)&  ($M_{\odot}$)& (kpc) & {($D_{Mpc}^{-1} \sin(i)$)}& {}\\
 & (1) & (2) & (3) & (4) & (5) & (6) & \\
\hline
Complexes & $3-15$ & 34$'${$^a$} &$\sim 10^{19}$ & $1-50 \times 10^5$ & $\lesssim 10$ & 34$'$ &  \citet{putman} \\
CHVCs{$^b$} & 0.78 & 2.7$'$ & $1.4 \times 10^{19}$ & $5\times10^4$ & 100 & 5.7$^{\circ}$ & \citet{2002AJ....123..873P}\\
 Disk-halo clouds & 0.030 & 6$''$ & $ 1 \times 10^{19}$ &  700 & 0.66 & 2.3$'$ & \citet{2010ApJ...722..367F}\\
\hline
\end{tabular}
{$^a$}{Typical assumed size $\sim 10$ kpc.} $^b${Assumed distance from
  galaxy center of 100 kpc.}\\ (1) diameter in kpc (2) angular diameter
normalized for distance (3) column density in cm$^{-2}$ (4) typical
\HI\ mass (5) height above the plane in kpc (6) projected separation
from the plane normalized for distance.\\
\end{table}

\section{Summary}

Two of the main unanswered questions in galaxy evolution are: ``how do
the astrophysics at the scales of individual clouds and gas complexes
lead to global star formation laws?'' and ``where do galaxies acquire
their gas from?'' Much progress has been made on these questions in
recent years, but the required combination of column density
sensitivity and high spatial resolution that is needed to
definitively answer these questions has so far been missing. For
example, a proper study of the first problem has so far really only
been possible in the MW and the nearest galaxies in the Local Group
(see the chapter by \citealt{naomi} elsewhere in this
volume). Similarly, the second question has so far only been addressed
for a limited number of galaxies with investment of a large amount of
observing time.  SKA will enable us to make great progress in both
areas.

For studies of individual objects, SKA1-MID provides the highest
resolution and sensitivity.  With a maximum effective angular
resolution of $3''$ SKA1-MID will go as deep as the highest-resolution
existing \HI\ surveys ($\sim 3 \cdot 10^{20}$ cm$^{-2}$ at 5$\sigma$
and 5 km s$^{-1}$ --- this limit definition applies to all values
quoted here) in the same observing time ($10^h$ per galaxy), but at a
resolution that is at least twice as good. SKA1-SUR with its 18 square
degree field of view will be able to instantaneously map the
environment around the target galaxies. While (for a fixed observing
time), its resolution and sensitivity will be more limited, this will
nevertheless provide invaluable information on the context of the
observed objects.

SKA2 will go an order of magnitude deeper than SKA1-MID and
SKA1-SUR. SKA2 will allow mapping of individual galaxies at
$1''$. After 10$^h$ it will reach column densities of $2.5 \cdot
10^{20}$ cm$^{-2}$, while 100$^h$ gives a $5\sigma$ limit of $8\cdot
10^{19}$. This allows mapping of galaxies out to 10 Mpc at a spatial
resolution that has so far only been possible in our Local Group.

SKA1-MID has an exquisite column density sensitivity, and at an
angular resolution of $30''$ is extremely efficient in detecting
low-column density \HI. At $10^h$ it will reach limits of $2.9\cdot
10^{18}$ cm$^{-2}$, which is twice as deep as the deepest existing
observations at comparable resolutions. A 100$^h$ integration gives
limits of $9.3 \cdot 10^{17}$ cm$^{-2}$, comparable to what is
achieved with the single-dish Green Bank Telescope, but at 20 times
better resolution. SKA1-SUR will again be able to provide quantitative
information on the environment and larger surroundings of any
detections.

SKA2 will be truly ground-breaking: after 10$^h$ its column
density limit at $30''$ angular resolution is $\sim 2.8 \cdot 10^{17}$
cm$^{-2}$, after 100$^h$, $\sim 8.9 \cdot 10^{16}$ cm$^{-2}$. The
latter will enable a direct exploration of the link between galaxies
and the cosmic web, and should reveal accretion processes in great
detail.

SKA will therefore revolutionize our understanding of star formation
in other galaxies, the processes leading to it, as well as the ways by
which galaxies acquire the necessary gas.

\bibliographystyle{apj}

\end{document}